\documentstyle[prb,aps,preprint]{revtex}

\begin{document}

\draft

\title{ Towards an understanding of CMR pyrochlore
Tl$_{2}$Mn$_{2}$O$_{7}$}

\author {C.I. Ventura~$^*$}

\address {Centro At\'omico Bariloche, 8400 - Bariloche, Argentina}

\author {M.A. Gusm\~ao}

\address {Instituto de F\'{\i}sica, Universidade Federal do Rio Grande do Sul,
CP 91501-970 Porto Alegre, RS, Brazil}

\maketitle

\begin{abstract}
The colossal magnetoresistance (CMR) exhibited by the pyrochlore
compound Tl$_{2}$Mn$_{2}$O$_{7}$ was discovered in 1996. Important
differences with CMR Mn-perovskite compounds regarding crystal
structure and electronic properties were soon realized, indicating the
possibility of different mechanisms for CMR being present.  Recently,
new pyrochlore compounds exhibiting CMR were prepared.  The larger
experimental background now existent, has received diverse
interpretations. In the present work, we aim to contribute to the
understanding of the properties of Tl$_{2}$Mn$_{2}$O$_{7}$ through the
study of a generic model for the compound, including Hund and
superexchange couplings as well as hybridization effects (HS model).
The model includes two kinds of electronic orbitals, as widely
believed to be present: one, directly related to the magnetism of the
compound, involves localized Mn moments, represented by spins, and a
narrow band strongly Hund-coupled to them. Also, more extended
electronic orbitals, related to the carriers in
Tl$_{2}$Mn$_{2}$O$_{7}$, appear and hybridize with the narrow band.
The possibility of a superexchange coupling between the localized
spins is also allowed for. This generic model allows exploration of
many of the proposals put forward by other researchers for
Tl$_{2}$Mn$_{2}$O$_{7}$. As a first approach to the problem, here we
employ a mean field approximation to study the magnetic phase-diagram
and electronic structure of the HS model, for various sets of
parameters mentioned before in connection with the
compound. Furthermore, we are able to exhibit similarities with
predictions obtained in 1997 using an intermediate valence model for
Tl$_{2}$Mn$_{2}$O$_{7}$, if appropriate sets of parameters are
used. In particular, we show here many common features which can be
obtained for the electronic structure around the Fermi level in the
ferromagnetic phase. Due to the simplifications adopted in the present
treatment the comparison between the models could not be extended to
higher temperatures here.
\end{abstract}

\vspace{2cm}



\narrowtext


\section{INTRODUCTION}
\label{sec:intro}

Since its discovery in 1996,\cite{5,6,7} the colossal
magnetoresistance (CMR) exhibited by the pyrochlore compound
Tl$_2$Mn$_2$O$_7$ posed many interesting questions. First among them,
whether a common fundamental mechanism could be responsible for the
origin of CMR both in this compound and in the larger family of
perovskite manganese oxides,\cite{3} which have different crystal
structure and electronic properties.  The recent appearance of CMR in
other compounds with the pyrochlore crystal structure
\cite{presion,shim99} exhibiting long range ferromagnetic ordering,
and the larger experimental background now existent, allow a better
standpoint to approach the theoretical description of this compound.

Tl$_2$Mn$_2$O$_7$ undergoes a ferromagnetic transition with critical
temperature\cite{6,shim99} $T_{c} \sim $ 125K (also reported\cite{5,7}
as 140K), below which temperature the compound exhibits long range
ferromagnetic ordering and is metallic. A sharp decrease in
resistivity with the development of spontaneous magnetization below
$T_{c}$, and the magnetoresistance maximum around $T_{c}$ with field
induced magnetization, which are similar to those of CMR
Mn-perovskites, indicate that a strong coupling between transport and
magnetism also exists in Tl$_2$Mn$_2$O$_7$.\cite{5,6,7} But important
differences have been established with respect to the CMR 
perovskites, starting from its A$_2$B$_2$O$_7$ ``pyrochlore'' crystal
structure.\cite{7'} No Mn$^{3+}$ has been detected,\cite{5,7,single}
so that the associated Jahn-Teller structural distortions\cite{4'} as
well as a possible double exchange (DE) mechanism\cite{2} inducing
ferromagnetism between Mn$^{3+}$-Mn$^{4+}$ pairs, studied in
connection with CMR in perovskite manganites,\cite{3,4'} are
absent. The crystal structure is almost temperature independent around
$T_{c}$, no structural anomalies appear which could be associated with
the change in magnetotransport properties, thus evidencing negligible
spin-lattice correlations.\cite{5} Structural analysis indicate
negligible O deficiencies present\cite{7} in stoichiometric
Tl$_2$Mn$_2$O$_7$, and recently specially synthesized
Tl$_{2-x}$Mn$_2$O$_{7-y}$ defect compounds show the development of
spin-glass behavior subjacent to ferromagnetism.\cite{Odefect}
Magnetization measurements for the stoichiometric compound have shown
saturation values of e.g. $\mu_{S} \sim$ 2.59,\cite{5}
2.7,\cite{7,rudoping} 2.8,\cite{scdoping} 3 $\mu_B$,\cite{shim99}
which are in general slightly lower than the 3$\mu_{B}$ expected for
Mn$^{4+}$ ions with a $3d^3$ ($S=3/2$) configuration.  Hall
experiments have indicated a very small number of high mobility
electron-like carriers: $\sim$0.001-0.005 conduction electrons per
formula unit,\cite{7,sushkolast} differing notably with the hole-doped
Mn-perovskites where CMR appears around 30$\%$ doping.  The critical
temperature $T_c$ has been reported to show a negative pressure
dependence,\cite{presion} again in contrast to CMR perovskites, though
above 10 Kbar a change of tendency and positive pressure shifts of
$T_c$ in Tl$_2$Mn$_2$O$_7$ have been reported.\cite{boston,dolores}
Recently, also striking differences with the perovskites were reported
for the evolution of the charge dynamics probed in optical
conductivity measurements.\cite{conduct} Experimental research of
magnetotransport has also been actively pursued in other pyrochlore
compounds where the ordered phase remains
insulating,\cite{presion,shim99} as well as in many obtained by
substitutional doping of the diverse components of
Tl$_2$Mn$_2$O$_7$.\cite{6,scdoping,rudoping,sbdoping,bidoping,Odefect,cddoping}

The experimental results have received various interpretations and
prompted different proposals and suggestions for the mechanism
potentially involved in CMR in Tl$_2$Mn$_2$O$_7$ since 1996.  Already
in Refs. 3 and 13 it was suggested that magnetism and transport seemed
related primarily to two different subsystems, unlike DE invoked for
CMR perovskites: magnetism (suggested to be possibly driven by
superexchange\cite{5,7}) involving mostly Mn-$3d$ and O orbitals,
while transport would be mainly determined by carriers involving
highly itinerant Tl-$6s$ orbitals mixed with oxygen.\cite{7} Unusually
large scattering from spin fluctuations accompanying the FM ordering
in the presence of low carrier densities was suggested as explanation
for CMR.\cite{7,scdoping} Electronic structure calculations\cite{bsub}
for Tl$_{2}$Mn$_{2}$O$_{7}$ indicated that the bands immediately below
the Fermi level correspond mainly to Mn-$t_{2g}$ states, above the
Fermi level appear Tl-$6s$ bands, while O states are mixed with these
around the Fermi level. The band structure calculation by Singh
\cite{bsingh} obtained a strong spin differentiation in the electronic
structure around the Fermi level in the ferromagnetic ground state.
The majority spin Mn-$t_{2g}-$O-$2p$ bands below the Fermi level are
separated by a gap from the Mn-$e_{g}$ derived bands, though there is
also a mixing with Tl and O states above and below the Fermi
level. Thus a small near band edge Fermi surface is obtained. For the
minority channel instead, a highly dispersive band with a strong
admixture of Mn, Tl and O was found around the Fermi level, leading to
a metallic minority spin channel.\cite{bsingh} In Ref. 3 it was also
mentioned that the Hall data could result from a small number of
carriers in the Tl-$6s$ band: assuming
Tl$_{2-x}^{3+}$Tl$_{x}^{2+}$Mn$_{2-x}^{4+}$Mn$_{x}^{5+}$O$_{7}$ with
x$\sim$ 0.005 the data could be accounted for, indicating a very small
doping into the Mn$^{4+}$ state.\cite{5,7} In 1997, this proposal was
explored in terms of an intermediate valence model (IVM) for
Tl$_{2}$Mn$_{2}$O$_{7}$ by Alascio and one of us
(C.I.V.).\cite{cec-blas} Hall experiments \cite{5} and CMR results
\cite{5,6,7} could be accounted for in the context of the strongly
differentiated electronic structure obtained,\cite{cec-blas} with
temperature and magnetic field changes which allowed explanations for
the drastic carrier density changes which had been observed at $T_{c}$
in Hall data,\cite{5} proposing the presence of a (spin-dependent)
hybridization gap or pseudogap near the Fermi level.  It may be
mentioned that while Mn$^{5+}$ is a rarely found oxidation state for
Mn, in Tl$_{2}$Mn$_{2}$O$_{7}$ it could effectively appear as a result
of the strong hybridization of orbitals near the Fermi
level. Furthermore, recently the presence of Mn$^{5+}$ ions has been
reported in Cd-substituted samples (Tl$_{2-x}$Cd$_{x}$Mn$_{2}$O$_{7}$)
together with Mn$^{4+}$ ions,\cite{cddoping} while in a different
context Mn$^{5+}$ in a tetrahedral environment has been found to exist
in Ba$_{3}$Mn$_{2}$O$_{8}$.\cite{mn5+}

In 1998, Majumdar and Littlewood \cite{littlewood} explored the
scenario of spin fluctuations around $T_{c}$ in presence of very low
carrier densities proposed in Refs.\ 3 and 13, starting from the
proposal of a microscopic model involving Hund and superexchange
coupling,\cite{littlewood} and showing that spin polarons might form
and provide an explanation for CMR.  An electronic structure
calculation by Mishra and Sathpathy\cite{mishra} in 1998 shows some
common features with the IVM band structure: in the FM phase they
mention the presence of a free-electron-like minority spin
$\Gamma_{1}$ band crossing the Fermi energy with highly mobile
carriers.  They relate this band to Mn-$t_{2g}$ and Tl-$6s$ orbitals
hybridized with O, the Mn-$e_{g}$ orbitals being higher in energy.
The occupied majority-spin bands consist of a Mn$-t_{2g}$/O band, and a
small pocket of very heavy holes, unimportant for
transport.\cite{mishra} They propose a ferromagnetic double exchange
Hamiltonian for Tl$_{2}$Mn$_{2}$O$_{7}$ with a very small number of
$\Gamma_{1}$ Zener carriers which align anti-parallel to the localized
Mn$-t_{2g}$ spins, including a weak AF-superexchange interaction
between the localized spins. Recent optical conductivity measurements
\cite{conduct} are consistent with the main features of the electronic
structure calculations mentioned.\cite{bsingh,cec-blas,mishra}
Recently, an explanation of the origin and pressure dependence of
ferromagnetism has been put forward in terms of a conventional FM
superexchange model,\cite{dolores} provided that the intra-atomic Hund
interactions in oxygen are also taken into account and the empty
Mn$-e_{g}$ orbitals hybridize with the O and Tl levels.

In this paper we aim to contribute to the understanding of CMR in
Tl$_2$Mn$_2$O$_7$ by presenting our investigation of the magnetic
phase diagram and electronic structure of a generic model for the
compound, which is based on the suggestions presented by many
researchers proposing the presence of superexchange interactions (of
either sign) between nearest neighbor 
Mn$^{4+}$ ions,\cite{5,7,littlewood,presion,dolores} and the
indications of previous band structure
calculations.\cite{bsub,bsingh,mishra} In Section \ref{sec:model} we
introduce this generic model for Tl$_2$Mn$_2$O$_7$, and make a brief
presentation of the mean-field treatment that we have employed in our
study.  Section \ref{sec:results} exhibits the results which we
obtained for different sets of parameters, including those that have
been suggested by various authors \cite{littlewood,mishra,dolores} as
possibly relevant for the description of this compound. A comparison
is also made with the electronic structure obtained before using the
IVM model,\cite{cec-blas} discussing the similar features which it has
been possible to exhibit between them.  In Section \ref{sec:summary} a
summary of our study of the HS model, and comments on the present
status of the understanding of the pyrochlore compound
Tl$_2$Mn$_2$O$_7$ are presented.

\section{ A MODEL FOR T\lowercase{l}$_2$M\lowercase{n}$_2$O$_7$ INCLUDING
HUND AND SUPEREXCHANGE COUPLINGS}
\label{sec:model}    

In this section we introduce a generic model for
Tl$_{2}$Mn$_{2}$O$_{7}$ based on the suggestions presented by many
researchers,\cite{5,7,presion,littlewood,mishra,dolores} general
enough to allow us to make a comparative study between the various
proposals which it includes as particular cases.  The generic model
takes into account the now widely accepted fact that in this material
transport and magnetism seem to be primarily related to different
electronic orbitals, \cite{7,scdoping,bidoping,rudoping,cddoping}
which we describe by two bands of different character which are
hybridized.  One band is mainly itinerant, and the other includes
strong correlations, as we now describe. The correlated electrons
appear forming a lattice of local magnetic moments, and also occupying
a narrow band, which are coupled via a strong Hund exchange
interaction. This would correspond to the assumption that the local
magnetic moments present in the material are due to Mn$^{4+}$ ions,
with occupied Mn-$t_{2g}$ orbitals which, however, due to their mixing
with O orbitals below the Fermi level, are not completely localized,
giving rise to the narrow band. A hybridization exists between the
correlated band and the mostly itinerant one, which is connected with
the low lying conduction states involving Tl$-6s$ orbitals and O
levels (eventually hybridized with the mostly empty Mn-$e_{g}$
orbitals\cite{dolores}) in accordance with most experimental
suggestions mentioned above, band structure
calculations,\cite{bsingh,mishra} and the recent description of the
pressure dependence of $T_{c}$ for various CMR
pyrochlores.\cite{dolores} The possibility of a superexchange coupling
between the localized spins being present is also
considered.\cite{5,7,littlewood,mishra,dolores}

 Concretely, the generic model that we have studied is described by 
the Hamiltonian 
\begin{equation}
H \,   =   \, H_{c} \, + \, H_{d} \, + \, H_{V}  \, + \, H_{S}
\, , \, 
\label{eq:H}
\end{equation}
where:
\begin{eqnarray}
H_c  &=&  - t_{c} \sum_{\langle i,j\rangle \sigma}   
(\, c^\dagger_{i\sigma} c^{}_{j\sigma}  \, + \,  \textrm{H.c.} \,)  - \, \mu   
\, \sum_{i\sigma} \, c^\dagger_{i\sigma} c^{}_{i\sigma} \; , \nonumber \\
H_d  &=&  - t_{d} \sum_{\langle i,j\rangle \sigma}   
(d^\dagger_{i\sigma} d^{}_{j\sigma}  +  \textrm{H.c.})  +
(\varepsilon_d - \mu)    
\sum_{i\sigma} d^\dagger_{i\sigma} d^{}_{i\sigma}  \; , \nonumber \\
H_{V}  &=&  V \sum_{i\sigma} \left(   
 c^\dagger_{i \sigma} d^{}_{i\sigma} + d^\dagger_{i\sigma} c^{}_{i\sigma}
  \right) \; , \nonumber \\
H_S  &=&  -  J_{\rm H}  \sum_i {\bf S}_i \cdot
{\bf s}^{(d)}_i - J_S \sum_{\langle i,j\rangle} {\bf S}_i \cdot
{\bf S}_j \; . 
\label{eq:Hparts} 
\end{eqnarray}

In this Hamiltonian $H_{c}$ describes the band associated to the
weakly (here simplified as un-) correlated carriers in
Tl$_{2}$Mn$_{2}$O$_{7}$, characterized by a hopping parameter $ t_{c}
$ between nearest neighbors, and a creation operator
$c^\dagger_{i\sigma}$ for an electron with spin $\sigma$ in the
respective Wannier orbital centered on site $i$.  $H_{d}$ describes
the narrow band associated to the correlated orbitals, characterized
by a hopping parameter $ t_{d} $, and a creation operator
$d^{\dagger}_{i\sigma}$ for an electron with spin $\sigma$ in the
Wannier orbital of reference energy $\varepsilon_{d}$ (with respect to
the center of the $c$ band) localized on site $i$. The chemical
potential of the system is denoted by $\mu$, and is determined by the
total electronic filling $n$.  The following term of the Hamiltonian
describes a hybridization, $H_{V}$, between the $c$ and $d$ electron
bands, which we assume to be local, for simplicity.

The last term of the Hamiltonian, $H_{S}$, describes the Hund (or
Kondo-like) coupling $J_{H}$ between the localized spins $ {\bf
{S}}_{i}$ (which would represent the spin operator associated to a
Mn$^{4+}$ ion at site $i$, for example) and the itinerant spins,
denoted by ${\bf {s}}^{(d)}_{i}$, corresponding to the narrow $d$ band
associated to the correlated electron orbitals related to Mn in
Tl$_{2}$Mn$_{2}$O$_{7}$.  A Heisenberg exchange term is also included
in the spin Hamiltonian $H_{S}$ to allow for a superexchange
interaction $J_{S}$ effective between the local moments of
nearest neighbor Mn ions.  The sign convention was chosen such that
any positive spin coupling constant describes a ferromagnetic (FM)
interaction, while negative ones will denote antiferromagnetic (AF)
coupling.

To study the magnetic phase diagram and electronic structure of the
 system described by the general Hamiltonian introduced above
 [hereafter to be denoted also as the Hund-superexchange (HS) model],
 we have employed the mean field approximation as briefly discussed
 below.

\subsection{Ferromagnetic phase} 
 
We investigated the existence of a uniform ferromagnetic phase,
introducing two order parameters to characterize it:
\begin{eqnarray}
 M  &= &  \langle  S^{z}_{i}  \rangle  \; ,  
\label{defM} \\
 m & = & \langle s^{(d)z}_{i} \rangle \, \equiv \, \frac{1}{2} \langle 
d^\dagger_{i \uparrow} d^{}_{i\uparrow} - d^\dagger_{i\downarrow} 
d^{}_{i\downarrow}
 \rangle \;,
\label{defm}
\end{eqnarray}
denoting by $ < O > $ the thermal (quantum) average of operator $O$  
in the system at temperature $T$ ($ T = 0$, for the system in its
ground state).

The corresponding mean-field Hamiltonian, after Fourier transforming
the band terms, takes the form
\widetext
\begin{eqnarray}
H_{F}^{\rm MF} & = & 
\sum_{{\bf k}\sigma} \left[ \tilde\epsilon_c ({\bf k}) n^{c}_{{\bf k}\sigma} 
+ \tilde\epsilon_{d\sigma} ({\bf k}) n^{d}_{{\bf k}\sigma} \right] 
+ V \sum_{{\bf k}\sigma} \left( c^\dagger_{{\bf k}\sigma} d^{}_{{\bf
 k}\sigma} + d^\dagger_{{\bf k}\sigma} c^{}_{{\bf k}\sigma} \right) 
\nonumber \\ & & \mbox{}
- \sum_i \left( J_{H} m + z J_{S} M \right) S^{z}_{i} \, 
+ \, J_{H} M m N \, + \, \frac{1}{2} N z J_{S} M^{2} \;,
\label{hfmfa}
\end{eqnarray}
\narrowtext \noindent where the Hund and Heisenberg exchange couplings
have been decoupled as usual in mean field approximation.  $N$ stands
for the total number of lattice sites, $z$ is the coordination number,
and we have defined
\begin{eqnarray}
\tilde\epsilon_c ({\bf k}) & \equiv & 
\epsilon^c_{\bf k} - \mu \;, \nonumber \\
\tilde\epsilon_{d\sigma} ({\bf k}) & \equiv &
\epsilon^{d}_{\bf k} -\mu - \sigma J_H M/2 \;,
\label{eq:epstilde}
\end{eqnarray}
in terms of the dispersion relations for the bare $c$ and $d$
bands, $\epsilon^c_{\bf k} = - t_{c} \sum_{\bbox{\delta}}
e^{-i {\bf k}\cdot \bbox{\delta}} $ and $ \epsilon^{d}_{\bf k} =
\varepsilon_d - t_d \sum_{\bbox{\delta}} e^{-i {\bf k}\cdot
\bbox{\delta}}$, where $\bbox{\delta}$ describes the vectors 
connecting any one lattice site with its $z$ nearest neighbors, 
and  $ n^{\alpha}_{k, \sigma} \equiv  \alpha^\dagger_{k, \sigma} 
 \alpha_{k, \sigma} \, $ , \, with \, $\alpha=
c,d \, ; \, \, \sigma = \uparrow, \downarrow $. 

The first two terms on the RHS of Eq.~(\ref{hfmfa}) describe two
hybridizing bands, one of them subject to an effective magnetic field
[through the product $J_H M$ appearing in $\tilde\epsilon_{d\sigma}
({\bf k})$, Eq.~(\ref{eq:epstilde})]. The third term corresponds to an
array of independent localized spins in the effective field
arising from their original mutual interaction as well as their
interaction with the $d$-band electrons. Finally, the last terms are
additive constants produced by the mean-field decoupling.

The band part of the mean-field Hamiltonian 
 $H_{F}^{\rm MF}$ is easily diagonalized, yielding the energy eigenvalues 
\begin{equation}
E^{\sigma}_{\pm}({\bf k})  =  \frac{1}{2} \left[
{\tilde\epsilon}_{c}({\bf k})  + 
{\tilde\epsilon}_{d\sigma}({\bf k})  
 \pm \sqrt{ [ {\tilde\epsilon}_{c}({\bf k}) 
- {\tilde\epsilon}_{d\sigma}({\bf k})]^{2} + 4 V^{2} } \right] ,  
\label{eq:esigma}
\end{equation}
which represent the ferromagnetic bands for spin $\sigma $.
The corresponding eigenvectors allow us to write the electron
operators of the $c$ and $d$ bands as combinations of operators
related to the new hybrid bands (and vice-versa), which is used in the
calculation of the relevant averages.

At a given temperature $T$, and for a fixed total band filling $n$,
the magnetization order parameters can be calculated self-consistently
through their definitions, Eqs.~(\ref{defM}) and (\ref{defm}), while
the chemical potential can be determined by the self-consistency
condition for the total number of electrons occupying the
ferromagnetic bands. We chose to rewrite this in terms 
 of the original $c$ and $d$ band fillings (denoted $<n_{c}>$
 and $<n_{d}>$, respectively):
\begin{equation}
n \,   =   \, <n_{c}> \, + \, <n_{d}> \, .
\end{equation}

Finally, the free energy per site of the ferromagnetic solution,
\begin{equation}
\label{eq:fe_tot}
F= F_B + F_S + F_0 \;,
\end{equation}
 is evaluated as the sum of three terms: the band free energy, the
localized spins contribution, and additive energy constants. These
terms are respectively given by
 \begin{eqnarray}
F_{B}  &  = &   -  T  \sum_{{\bf k}\sigma} \sum_{\alpha=\pm} 
\ln \left[ 1 + e^{-  E^{\sigma}_{\alpha}({\bf k})/  T} \right]   
, \label{eq:FB} \\ 
F_{S}  &  =  & -  T  \ln {\rm Tr} \left[ e^{(J_{H} m + z J_{S} M )
S^{z} /  T}  \right] , \label{eq:FS} \\
F_{0}  &  =  &    J_{H}  M  m   +  \frac{1}{2} z J_{S}  M^{2} + \mu n
\;.  \label{eq:F0}
\end{eqnarray}
The last term in Eq.~(\ref{eq:F0}) must be added to compensate for the
inclusion of the chemical potential in the Hamiltonian defined
previously. Notice that we
are dropping the Boltzmann constant, which implies that temperature
will be expressed in the same units as energy.

\subsection{Antiferromagnetic phase}
 
We also investigated the existence of an antiferromagnetic phase.
Such long-range magnetic ordering would appear frustrated in the  
real pyrochlore lattice, but as a first approach, 
we study here a simplified version of the problem on a cubic lattice. 
In this case, the lattice is divided into two interpenetrating 
equivalent sublattices with opposite magnetizations, 
labeled $A$ and $B$ respectively. By symmetry, 
only two order parameters which describe the 
sublattice magnetizations characterize this phase:

\begin{eqnarray}
 \langle  S^{z }_{i}  \rangle   &  =  &  \left\{ 
\begin{array}{rl}
 M  , &  \quad  i \in A, \\
 -M , &  \quad  i \in B,
\end{array}
\right.
\label{defMAF} \\
 \langle  s^{(d)z}_{i}  \rangle & =  &   \left\{ 
\begin{array}{rl}
 m  , &  \quad  i \in A, \\
 -m , &  \quad  i \in B,
\end{array}
\right.
\label{defmAF}
\end{eqnarray}
where both $M$ and $m$ are positive-defined quantities.

The procedure used to determine the antiferromagnetic solution in mean
field approximation is similar to the one presented above for the
ferromagnetic phase. For completeness, we will only include here the
form assumed by the mean-field Hamiltonian, $H^{\rm MF}_{\rm AF}$,
written in terms of operators characterizing the two interpenetrating
sublattices: 
\widetext
\begin{eqnarray}
H^{\rm MF}_{\rm AF} &  =  & 
- \mu \sum_{{\bf k}\sigma} \left( a^\dagger_{{\bf k}\sigma} a^{}_{{\bf k}\sigma}  
+ b^\dagger_{{\bf k}\sigma} b^{}_{{\bf k}\sigma}  \right)
  + \sum_{{\bf k}\sigma} \epsilon^c_{\bf k}
 \left(    e^{- i {\bf k} \cdot {\bbox{\delta}}_0} 
a^\dagger_{{\bf k}\sigma} b^{}_{{\bf k}\sigma}  + \mbox{H.c.}
\right)    \nonumber \\  &  &   \mbox{}
+   \sum_{{\bf k}\sigma} \left[  \left( \varepsilon_{d} - \mu  -
 \sigma  J_{H}   M /2  \right)  A^\dagger_{{\bf k}\sigma} A^{}_{{\bf k}\sigma} 
 +    \left(  \varepsilon_{d} - \mu  + 
 \sigma  J_{H}   M/2  \right)  B^\dagger_{{\bf k}\sigma}
B^{}_{{\bf k}\sigma}  \right]
   \nonumber \\   &  &  \mbox{}
+ \sum_{{\bf k}\sigma} ( \epsilon^d_{\bf k}  -
\varepsilon_{d})   
\left(     e^{- i {\bf k} \cdot {\bbox{\delta}}_0}   
A^\dagger_{{\bf k}\sigma} B^{}_{{\bf k}\sigma}  + \mbox{H.c.} \right)  
+   V   \sum_{{\bf k}\sigma} \left(   
 a^\dagger_{{\bf k}\sigma} A^{}_{{\bf k}\sigma} + b^\dagger_{{\bf
k}\sigma} B^{}_{{\bf k}\sigma} + \mbox{H.c.}  \right)  
 \nonumber \\ &    &  \mbox{} 
- \sum_{ i = 1 }^{N/2} 
\left(  J_{H}  m  - z  J_{S}  M  \right) \left[ S^{z(A)}_{i} 
  -  S^{z(B)}_{i} \right]
+   J_{H}  M  m  N  +  \frac{1}{2} z J_{S}  M^{2} N \;. 
\label{afmfa}
\end{eqnarray}
\narrowtext\noindent Here we have introduced two-site unit cells to
describe the AF lattice and used the following notation: $ d_{i\sigma}
= A_{i\sigma} ( B_{i\sigma})$ for $i \in A (B) $; $ c_{i\sigma} =
a_{i\sigma} ( b_{i\sigma})$ for $i \in A (B) $; explicit indication of
sublattice is made for localized-spin operators; $ {\bbox{\delta}}_{0}
$ is a vector connecting the two sites (one belonging to each
sublattice) inside a unit cell. Notice that the wave-vector sums are
now performed over the reduced Brillouin zone of the AF lattice.
 
The determination of the self-consistent AF solution follows 
 along the same lines as in the FM case,
except that we have to obtain numerically the eigenvalues and
eigenvectors of the itinerant part of the effective Hamiltonian
(\ref{afmfa}), as we are now dealing with a 4x4 matrix.

\section{RESULTS AND DISCUSSION}
\label{sec:results}
 
In this section we present some of the results we have obtained for
the generic HS model introduced in the previous section using many
different sets of parameters: including those which have been
suggested by various authors \cite{bsingh,littlewood,mishra,dolores}
as possibly relevant for the description of
Tl$_{2}$Mn$_{2}$O$_{7}$. We want to stress that even with the
simplifications adopted in the present treatment, notwithstanding the
known over-estimation of critical temperatures and other limitations
to be mentioned, which are inherent to the mean-field approximation
used, it is nevertheless interesting to be able to exhibit features in
common with the different models that have been proposed for
Tl$_{2}$Mn$_{2}$O$_{7}$ if appropriate relevant sets of parameters are
chosen. This will be the aim of the presentation of results and
discussion in this section.

For simplicity and without much loss of generality, we assume the
localized spins to have $ S = 1/2$. In this case, the mean-field
self-consistency equations for the magnetization order parameter $M$
[Eqs.~(\ref{defM}) and (\ref{defMAF}), respectively], assume the
simple form
\begin{equation}
M  \,  =  \, \frac{1}{2} \, \tanh \left[ \frac{1}{2 
T} \left(  \, J_{H} \, m \, \pm  
\,z \, J_{S} \, M \, \right)  \right] \, ,   
\label{eq:weissms} 
\end{equation}
where the upper (lower) sign corresponds to the ferromagnetic
(antiferromagnetic) solution.

We have assumed a three-dimensional simple cubic lattice ($ z = 6 $),
and used a semi-elliptic approximation for the bare densities of states
(DOS). The $c$ band half-bandwidth $ W = z t_{c} $ is used as the
energy unit. To fix ideas in the following discussion we will choose
that value to be $\simeq 1 \;e$V, in qualitative agreement with
experiments and band-structure calculations.\cite{bsingh,mishra} We
also consider anti-homotetic bands, i.e., $ t_{d} \, = \ - \alpha \,
t_{c} $, $ \alpha$ being a (positive) narrowing factor.  This would be
consistent with an electron-like $c$ band related to the mobile
carriers,\cite{5,bsingh,mishra} and a narrower hole-like $d$
band.\cite{mishra,bsingh}

We have investigated the existence of magnetically ordered solutions
for many sets of parameters of the HS model.  Both FM and AF solutions
are obtained at low temperatures for wide ranges of parameter
values. This puts into evidence a strong competition between these two
kinds of ordering in the model. We select the physically relevant
solution in each case by comparing the free energies and choosing the
lowest one. In this paper we present ground-state phase diagram
results (at temperatures much lower than the critical ones for either
magnetic phase, $ T << T_{c_{(F)}},\,T_{c_{(AF)}} $, where full spin
polarization has been established) for which the bare $d$ band has
been centered at an energy $ \varepsilon_{d} \, = \, -0.7 \;e$V \, (the
bare $c$ band being centered at the origin). This is consistent with
the $d$ band lying near the bottom of the $c$ band, which will yield a
small number of conduction electrons.  Our phase diagrams are plotted
as functions of the bandwidth ratio $ \alpha$ and the
superexchange-coupling strength $J_{S}$. In each phase diagram the
following remaining parameters have been fixed: the Hund-coupling
strength $J_{H}$ (kept at a large value, comparable to the width of
the conduction band), the hybridization $V$ and the total electron
filling $n$.

A fact worth noticing is a symmetry manifest in the mean-field
 Hamiltonians (Eqs.~(\ref{hfmfa}) and (\ref{afmfa}), respectively):
 the system is invariant under a simultaneous change of sign of the
 Hund-coupling $J_{H}$ and of the relative orientation between
 magnetizations $m$ and $M$.  Thus, the system will not distinguish
 energetically between a situation corresponding to a FM-like
 Hund-coupling, in which case the magnetizations $M$ and $m$ will be
 parallel, and the case of AF-like Hund coupling, in which these two
 magnetizations would have relative anti-parallel orientations.
 Taking this symmetry into account, one can restrict the analysis to
 positive $J_{H}$, as expected on physical grounds for this compound.

We present phase-diagrams obtained for a range of electron fillings 
around $ n = 1$, near the region of interest for the description 
of  Tl$_{2}$Mn$_{2}$O$_{7}$ as will be discussed shortly in connection
with the electronic structure and occupation of the bands.   

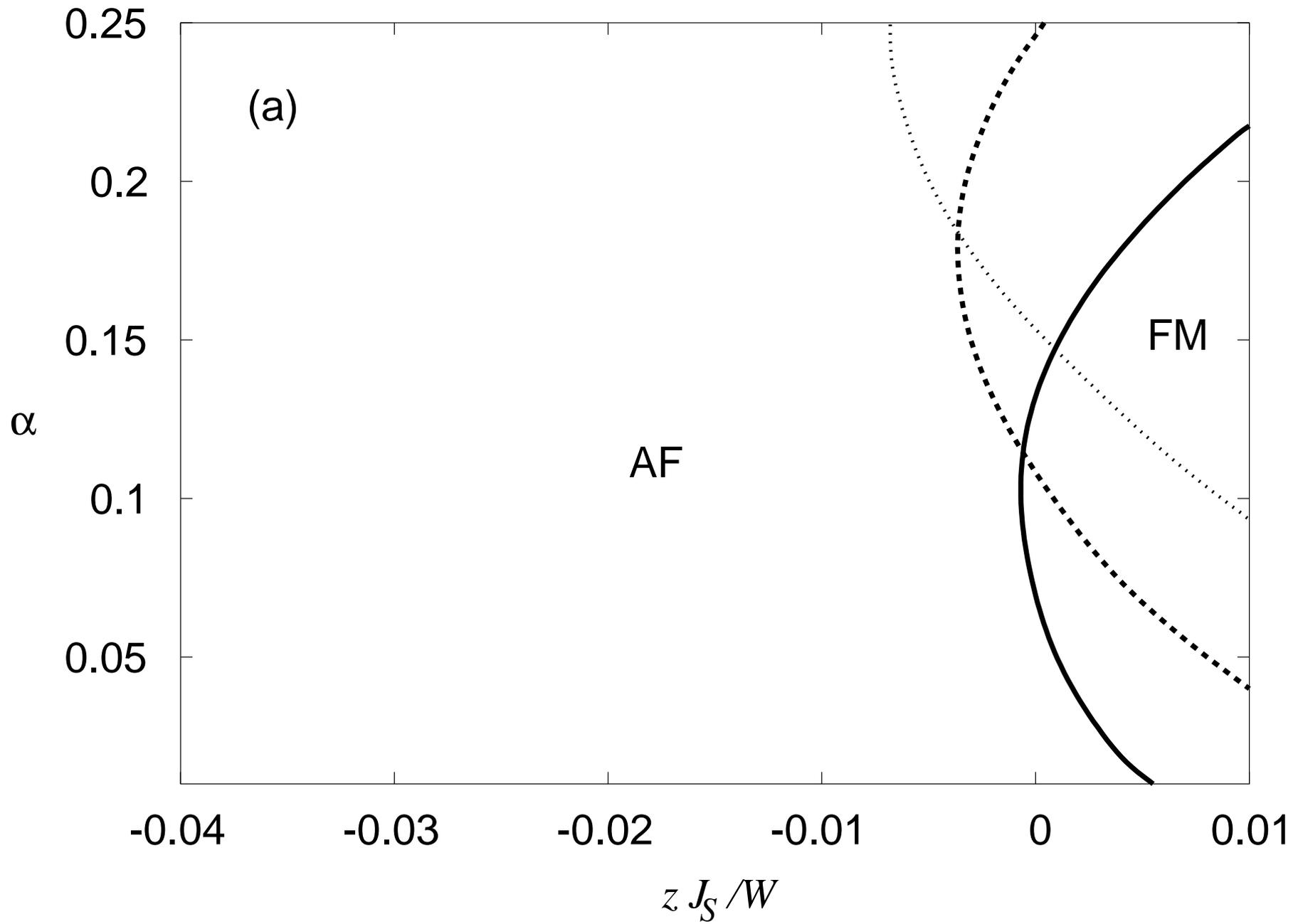
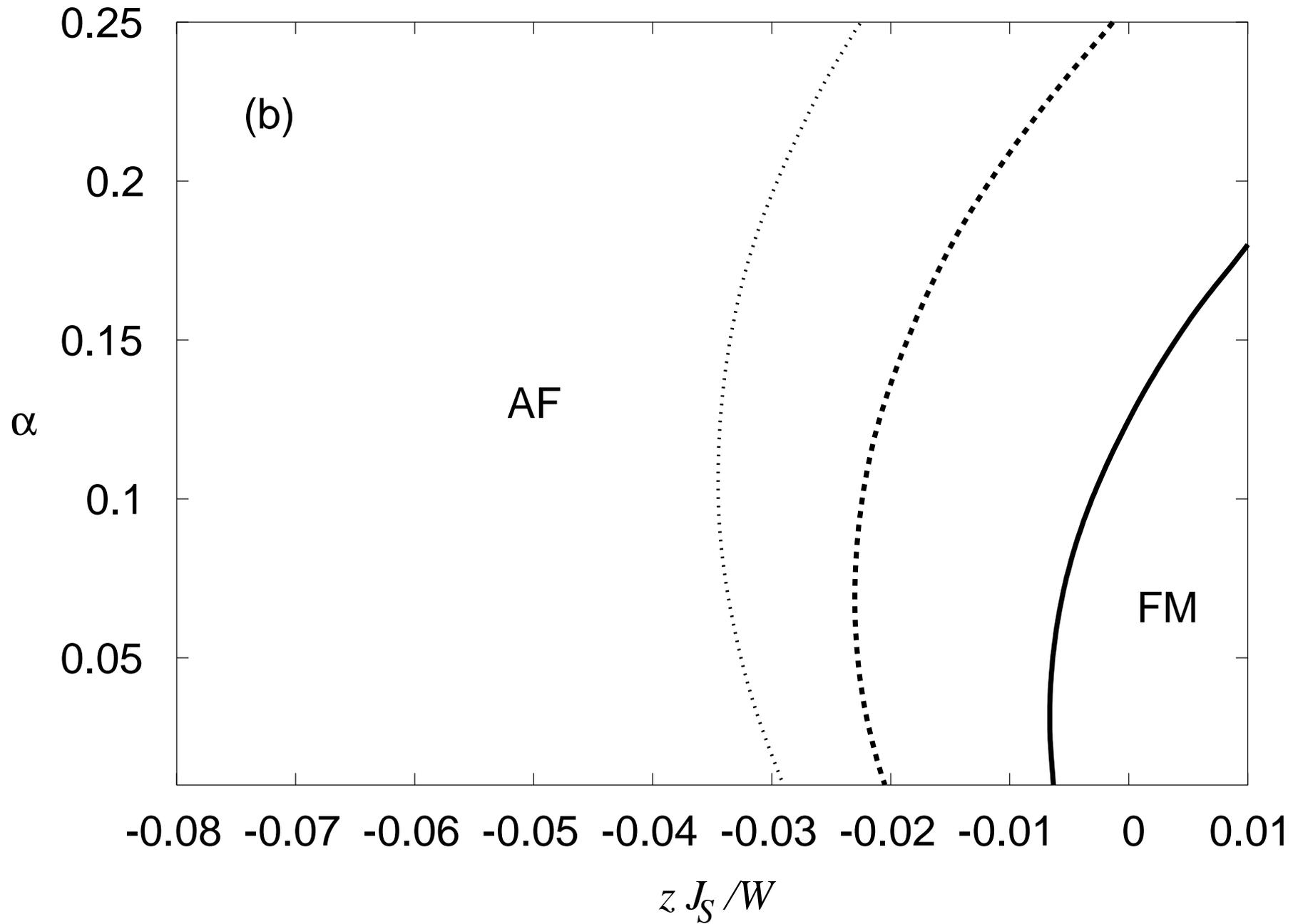
\begin{figure}
\caption{ Zero-temperature phase diagram of the HS model as a
function of bandwidth ratio $\alpha$, and superexchange coupling $ z
J_{S}$. (FM): ferromagnetic phase; (AF): antiferromagnetic phase.  
Parameters:  $ W \, = \, z \, t_{c} \, = \, 1\; e$V; 
$\varepsilon_d = - 0.7 \;e$V; $ J_{H} =  1.5 \;e$V. 
Phase transition lines shown for hybridizations:  $ V = 0.2 \;e$V
(solid line),  $  0.3 \;e$V (dashed line), and $ 0.4 \;e$V (dotted line).
(a) $ n = 1 $; (b) $ n = 1.09 $. }
\label{fig:pdn1}
\end{figure}

First, we will discuss the phase-diagrams obtained including an
antiferromagnetic superexchange (SE) coupling between nearest
neighbor spins, as some authors\cite{presion,mishra} have suggested: 
$ J_{S} \leq 0 $.  We can make a few general statements:

i) for small enough hybridization (e.g. $V = 0.02\; e$V ) no
ferromagnetism is obtained for any finite value of AF-superexchange in
the large region of variation of the bandwidth ratio $\alpha$
investigated ($ 0 \leq \alpha \leq 0.5 $ ) for fillings around $n =
1$;

ii) at larger hybridization values (e.g., $ V = 0.2, 0.3, 0.4\; e$V )
an interesting competition between AF and F is obtained as a function
of the parameters, including reentrant behavior as a function of
$\alpha$.  Though antiferromagnetism widely dominates for $ n= 1$, we
also found that for weak AF-superexchange, $ -0.01 eV \leq z J_{S}
\leq 0$, pockets of ferromagnetism appear at small to intermediate
values of $\alpha$. In Fig.~\ref{fig:pdn1}(a) we show how these FM
pockets shift to larger $\alpha$ values if the hybridization is
increased.

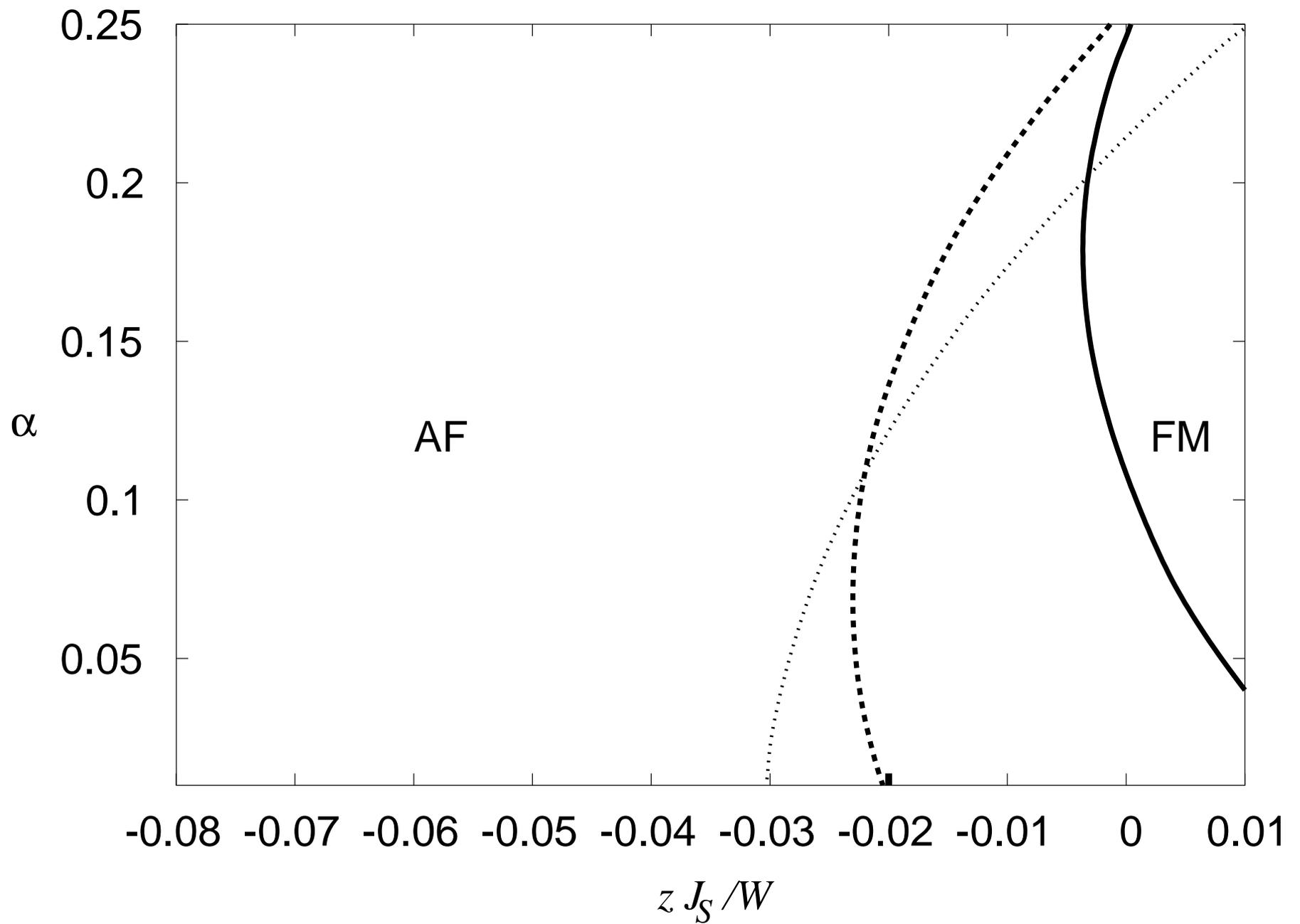
\begin{figure}
\caption{ Zero-temperature phase diagram of the HS model: dependence
on filling.  Same notation and parameters as in Fig.~\ref{fig:pdn1},
except for: $ V = 0.3 \;e$V; $ n = 1$ (solid line), $ 1.09 $ (dashed
line), $ 1.15 $ (dotted line).}
\label{fig:pdv03}
\end{figure}

  Larger electron fillings [see Fig.~\ref{fig:pdn1}(b)] increase the
size of the ferromagnetic regions, which are now obtained also at
smaller $\alpha$ values (narrower bare $d$ band).  To exemplify the
dependence of the phase diagram with electron filling at fixed
hybridization we include Fig.~\ref{fig:pdv03} corresponding to
hybridization $V = 0.3 \;e$V, where the cases $ n = 1, 1.09, 1.15 $
are shown.  The effect of filling on the phase diagram is amplified
for larger hybridization values.

Thus, varying the filling, in the presence of a hybridization large
enough, and for a $d$ band relatively narrower than the $c$ band, we
have demonstrated that ferromagnetism can be obtained from the HS
model even in the presence of a (weak) AF-superexchange($ J_{S} \leq
0$), as has been suggested before,\cite{mishra} the stability region
growing with increasing hybridization.

If the phase diagram in the presence of a weak FM superexchange
interaction ($ J_{S} > 0$) is now investigated, as proposed early on
for this compound by many researchers,\cite{5,6,7,bsingh,littlewood}
and later claimed to be effectively the case if a careful
consideration of the effect of O-orbitals was
undertaken,\cite{dolores} of course the ferromagnetic phase is
favored.  Now much larger ranges of the bandwidth ratio $\alpha$
correspond to ferromagnetic ground states, as can be appreciated in
Figs.~\ref{fig:pdn1} and \ref{fig:pdv03}.

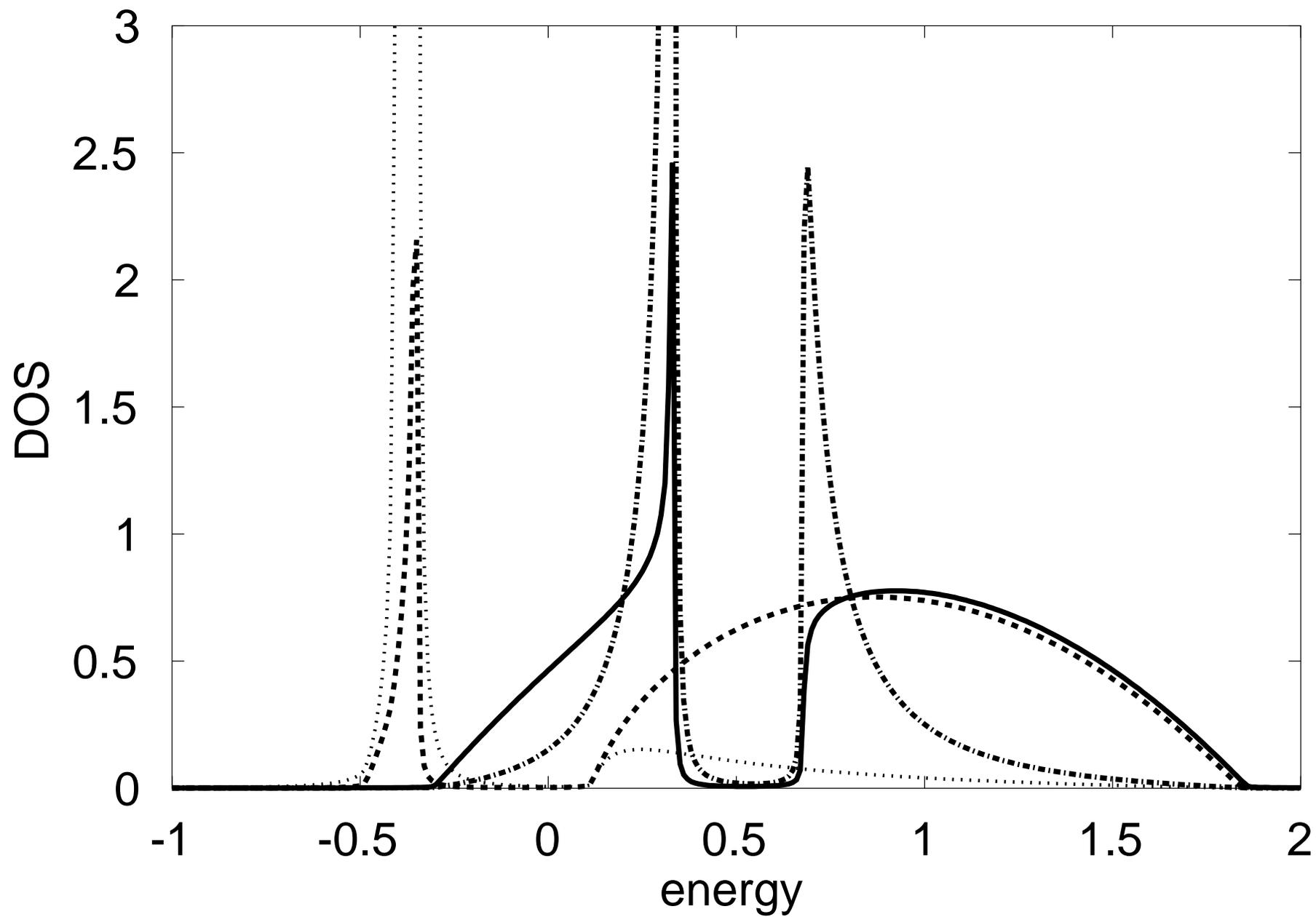
\begin{figure}
\caption{ Ferromagnetic phase of HS model: density of states as a
function of energy (measured with respect to the Fermi level, in units
of $W$), projected on the original bands.  $\rho_{d,\uparrow}$ (dotted
line), $\rho_{d\downarrow}$ (dot-dashed line), $\rho_{c\uparrow}$
(dashed line), $\rho_{c\downarrow}$ (solid line), for $ V = 0.3 \;e$V,
$ n = 1.09 $, $ \alpha = 0.1 $, $ z J_{S} = -0.01 \;e$V.  Other
parameters as in Fig.~\ref{fig:pdn1}.}
\label{fig:fmbands}
\end{figure}

The subtle interplay of the different parameters which has been shown
to determine the magnetic ordering in the HS model, in particular
needed to obtain a ferromagnetic ground state for example, can be
quite easily understood if one analyzes the corresponding electronic
band structure.  As seen in Fig.~\ref{fig:fmbands}, where we plot the
density of states decomposed in terms of the original bands, we find
that for the parameter ranges in which ferromagnetism is clearly
stable for the HS model, strong spin differentiation exists in the
electronic structure.  A band gap connected to the majority-spin
orbitals is present at the Fermi level ($E_{F}$), 
while a finite density of states  at $E_{F}$ 
is obtained mainly for the minority-spin  conduction 
band $c_\downarrow$.  
For the fillings which have been discussed above, the chemical
potential of the system is located near a hybridization band gap edge
(see Fig.~\ref{fig:fmbands}), providing occupation of the different
orbitals in qualitative agreement with previous electronic structure
calculations.\cite{bsingh,mishra} These results are also interesting
as they place us in an analogous scenario to that proposed in 1997 for
this compound,\cite{cec-blas} when the IVM for Tl$_{2}$Mn$_{2}$O$_{7}$
was introduced and shown to provide an explanation for the results
obtained in Hall experiments and the colossal nature of the MR of the
compound. The presence of the Fermi level near band edges, e.g. of a
hybridization gap or pseudogap, in a strongly spin differentiated
electronic band structure, and especially the spin-dependent
conduction band changes predicted as a function of temperature and
magnetic field, were shown\cite{cec-blas} to originate drastic changes
in the carrier concentration around $T_{c}$ which, in the absence of
anomalous Hall contributions, could be directly related to the
measurements in Hall experiments \cite{5} and explain them. The
validity of the Hall data analysis performed in Ref.\ 23 has recently
been confirmed by new experimental reports on Hall and thermopower
data over wider temperature and magnetic field
ranges.\cite{sushkolast} In Ref. 14 the new Hall data are successfully
explained using the same approach of Ref. 23, and support the
prediction of drastic carrier density changes near $T_{c}$, related to
conduction band changes due to the development of magnetization, as
being the main sources for the observed magnetotransport
properties.\cite{cec-blas,sushkolast,conduct}

Due to the simple mean-field approximation adopted, only at low
temperatures, when the system is fully polarized, are our phase
diagram results trustful, and the electronic structure results
suitable for comparison with those obtained before using the
IVM.\cite{cec-blas} The effect of spin-polarization by the Hund
interaction is not correctly estimated here at higher temperatures,
being included only through the mean field magnetization order
parameters.  In fact, an investigation of the critical temperatures
obtained here also evidenced problems related to the inability of the
MFA employed to treat adequately the strong Hund correlations: the
Curie temperatures obtained are found to decrease as hybridization is
increased (which one would tend to connect with negative pressure
coefficients for $T_{c}$ as experimentally observed
\cite{presion,boston,dolores}) even in cases for which the phase
diagrams show a stabilization of ferromagnetic ordering with increase
of the hybridization (as we obtain for fillings $ n > 1 $).  This
seems caused by a spurious tendency to stabilize a local magnetization
due to the mutual interaction between the localized spins and the $d$
electrons through their Hund coupling, irrespective of the intersite
interactions, due to the mean field approximation employed.  Thus, the
shortcomings of the present treatment for all but the lowest
temperatures prevent us from addressing here the interesting issue of
the ability of the HS model to describe the effect of pressure on
$T_{c}$ for different values of its parameters.  The increase of spin
disorder with temperature, and in particular the paramagnetic phase,
should be much better described using the coherent potential
approximation as in Ref.\ 23 for the IVM. We expect that such a
treatment for the paramagnetic phase of the Hund-superexchange model
would allow us to exhibit similarities with the IVM electronic
structure results above $T_{c}$, like those shown to exist for the
ferromagnetic phase in the present work, and to address other issues
of interest outside the scope of the simple treatment presented in
this work.

\section{SUMMARY}
\label{sec:summary}      

Increasing experimental evidence on magnetotransport in
Tl$_{2}$Mn$_{2}$O$_{7}$ points towards drastic changes in the carrier
concentration around $T_{c}$ as being mainly responsible for its
colossal magnetoresistance.  The absence of anomalous contributions to
the Hall resistance has also recently been confirmed.\cite{sushkolast}
Since the nature of the carriers involved in magnetotransport has been
established as free electron-like, there are now strong indications in
favor of a direct relationship between drastic changes in the
conduction band structure around $T_{c}$ and the observed
magnetotransport properties.

A strongly spin-differentiated electronic band structure, with drastic
changes in the conduction bands around the critical temperature, 
seems a natural way to explain such abrupt carrier concentration changes.   
The presence of hybridization gaps or pseudogaps which, depending on
the spin orientation of the carriers, might change
drastically at $T_{c}$ would allow for such a scenario. 

The present paper is a first approach to show how such a scenario,
first proposed for Tl$_{2}$Mn$_{2}$O$_{7}$ in 1997 in connection with
an intermediate valence model (IVM) introduced to describe CMR and
Hall data of the compound,\cite{cec-blas} can effectively also be
realized starting from different microscopic models suggested for
Tl$_{2}$Mn$_{2}$O$_{7}$, involving Hund and superexchange magnetic
couplings, as well as hybridization
effects.\cite{5,6,7,bsingh,littlewood,mishra,dolores} We have
performed a detailed study of the stability of magnetically ordered
phases obtained at low temperatures for various sets of parameters of
a generic Hund-superexchange model (HS). The model is based on the
assumption that the Mn orbitals, besides providing localized magnetic
moments, also show some degree of delocalization through hybridization
with the other orbitals, thus yielding a narrow band strongly coupled
to the localized moments via a Hund coupling. This narrow band also
shows strong hybridization with a broader conduction band, related to
the carriers in the compound. The possibility of a superexchange
interaction between the localized spins is also allowed for in the
generic model, thereby including many of the proposals presented
before for this compound.  The electronic structure of the HS model
was also investigated, obtaining for the ferromagnetic case a strong
spin differentiation and the presence of a minority-spin metallic
channel, thus exhibiting many similarities with the electronic
structure derived from the IVM  for this ordered phase.  The
simple mean-field approach adopted here is unsuitable to describe
adequately the effects of the Hund interaction as temperature is
increased. This limits the validity of our results to temperatures
well below the critical one. It also prevents us from discussing the
ability of the Hund-superexchange model to describe the pressure
dependence\cite{presion,boston,dolores} of $ T_{c}$ in its various
parameter regimes.

Nowadays the mechanism responsible for the long-range ferromagnetism
experimentally observed in Tl$_2$Mn$_2$O$_7$ has not yet been
conclusively identified. The results presented here, those obtained
before with the IVM,\cite{cec-blas} as well as the conclusions of much
recent experimental work mentioned here, seem to indicate that in fact,
provided the transport remains metallic, the detailed model for the
magnetism matters much less than the trend with carrier density, as
has been pointed out before for Tl$_{2}$Mn$_{2}$O$_{7}$.\cite{pl2} In
fact, our magnetic phase diagram results show that, although a
ferromagnetic superexchange interaction between the Mn localized
moments might be necessary to explain ferromagnetism in the related
insulating pyrochlores (for example, where Y or Lu appear substituting
Tl), the enhanced stability of FM ordering in Tl$_{2}$Mn$_{2}$O$_{7}$
can be connected to the presence of spin-polarized charge carriers and
hybridization effects, in accordance with previous
suggestions.\cite{dolores,shim99} Careful analysis of spectroscopic
experiments is needed to elucidate conclusively whether the drastic
changes observed in carrier concentration are mainly related to
magnetization-related changes of the conduction bands themselves
and/or shifts of the conduction band edges.\cite{sushkolast,conduct}
  
\acknowledgments 

We especially thank Blas Alascio for enlightening discussions, and
acknowledge financial support from the Argentina-Brazil cooperation
program SECYT-CAPES BR/A99-EIII18/00. M.A.G. was also supported by
grant PRONEX/FINEP/CNPq 41.96.0907.00 (Brazil).

\end{document}